\documentclass{ifacconf}

\usepackage{graphicx}      
\usepackage{natbib}        
\usepackage{amsmath}
\newtheorem{Problem}{Problem}
\begin{document}
\begin{frontmatter}

\title{CBF-based Driving Assistance for\\ Traffic Flow Stabilization\thanksref{footnoteinfo}} 

\thanks[footnoteinfo]{This work is supported by Mazda Motor Corporation.}

\author[First]{H.~Irie}\author[First]{M.~Inoue}\author[First]{B.~Okita}
\author[Second]{A.~Yamaguchi}\author[Second]{T.~Taki}\author[Second]{T.~Hatano}

\address[First]{Keio University, Kanagawa, Japan
(e-mail: irihaya0913@keio.jp, minoue.z6@keio.jp, banri.okita@keio.jp).}
\address[Second]{Mazda Motor Corporation, Hiroshima, Japan
(e-mail: yamaguchi.akira@mazda.co.jp, taki.t@mazda.co.jp, hatano.taka@mazda.co.jp)}

\begin{abstract}                
This manuscript addresses a hierarchical control system designed to suppress traffic congestion. The lower-layered controllers, implemented in each controlled vehicle, monitor microscopic vehicle behaviors and assist human drivers to ensure sufficient spacing for following vehicles. This spacing logic is designed based on the Control Barrier Function. Meanwhile, the upper-layered controller monitors the macroscopic traffic flow and activates the necessary lower-layered controllers, using a data-driven approach for the activation logic design.
Furthermore, the effectiveness of the proposed control system is evaluated in a traffic flow simulation environment constructed using real-world traffic data.

\end{abstract}


\end{frontmatter}

\section{Introduction}

In recent years, automotive technologies have significantly advanced, and research and development have mainly aimed at improving performance indices of individual vehicles, such as fuel efficiency, ride comfort, and safety (see e.g., the work by \cite{7490340}).  
While these improved driving behaviors tend to prioritize individual vehicle performance, they often compromise the performance of surrounding vehicles and overall macroscopic traffic stability. 
(see e.g., the works by \cite{531196,6515636, Sugiyama_2008}).
From this perspective, in addition to conventional performance indices, cooperative driving factors must be taken into account to suppress traffic congestion by using information on surrounding vehicles and macroscopic traffic conditions (see e.g., the works by \cite{9037275,TALEBPOUR2016143}).
 In particular, appropriate velocity control according to macroscopic traffic conditions is expected to suppress velocity fluctuations of following vehicles and contribute to traffic flow stabilization.

Based on this background, many studies have focused on vehicle control for traffic flow stabilization.
\cite{Zhao_2023,li2025safety} proposed a safety filter based on Control Barrier Function (CBF) (see e.g., the work by \cite{8796030}), which modifies the control input generated by a velocity controller for traffic flow stabilization. This approach achieves both traffic flow stabilization and vehicle safety.
In addition, as a study on cooperative vehicle control using macroscopic traffic conditions, \cite{9928340,FAN2025100188} proposed a hierarchical control system. The upper-layered controller monitors macroscopic traffic conditions such as traffic density and generates control commands aimed at improving the overall efficiency of traffic flow. Meanwhile, the lower-layered controller regulates the microscopic behaviors of individual vehicles to ensure safety and traffic stability.
The proposed systems assume direct control of connected and automated vehicle platoons through vehicle-to-vehicle communication, enabling cooperative coordination among the vehicles in the platoon.

\begin{figure}[t]
\centering
\includegraphics[width=0.95\linewidth]{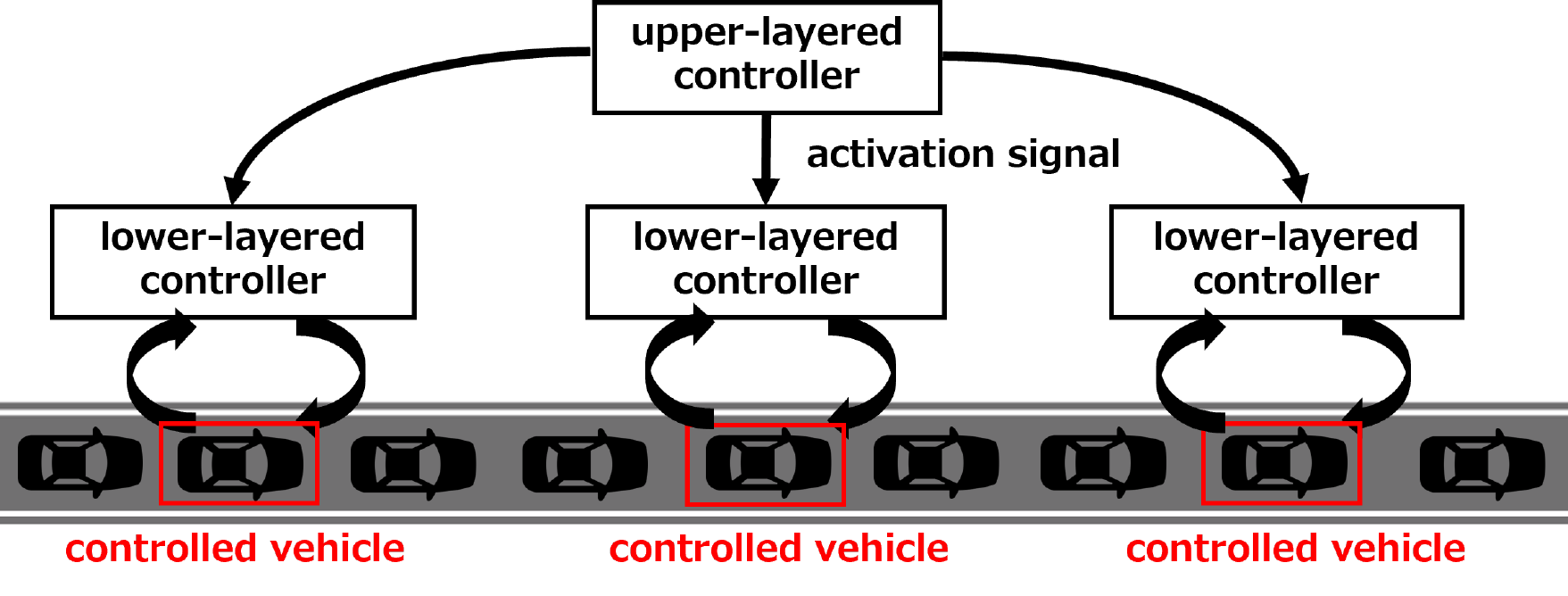}
\caption{Hierarchical control system for traffic flow stabiliztion}
\label{fig}
\end{figure}

This manuscript addresses the stabilization of traffic flow by controlling microscopic vehicle behaviors based on macroscopic traffic conditions. Specifically, rather than assuming fully autonomous driving, we design advanced driver assistance systems (ADAS) tailored for human-driven vehicles.
In ADAS, excessively frequent interventions may cause annoyance and discomfort to drivers, leading to reduced system acceptability (see e.g., the work by \cite{su13073932}). Therefore, unnecessary interventions should be suppressed while maintaining safety and traffic flow stabilization.
We propose a hierarchical control system to suppress traffic congestion while reducing unnecessary interventions. A conceptual diagram of the proposed system is shown in Fig.~\ref{fig}.
Lower-layered controllers, implemented in each controlled vehicle, monitor microscopic vehicle behaviors and intervene in driver control actions based on CBF to ensure sufficient spacing for following vehicles. The upper-layered controller monitors macroscopic traffic flow and generates an activation signal based on an intervention logic designed through a data-driven approach. The activation signal is then used to activate the necessary lower-layered controllers.

\section{Hierarchical Control System}

\subsection{Overview}

This manuscript addresses a control problem in a traffic scenario where multiple vehicles travel in a platoon on a single-lane road, as shown in Fig.~\ref{fig}.

Our objective is to suppress traffic congestion in the platoon.  
Traffic congestion is generated and spreads through the accumulation of microscopic car-following behaviors, in which each vehicle adjusts its motion according to the inter-vehicle distance and relative velocity with respect to the preceding vehicle (see e.g., the works by \cite{531196,6515636,Sugiyama_2008}).  
Therefore, the solution to traffic congestion lies in the appropriate control of car-following behavior.

However, applying control interventions to all vehicles in a real-world traffic system is impractical. Therefore, the proposed system selectively intervenes in only a relatively small number of controlled vehicles. In addition, excessively frequent interventions may increase driver burden and reduce system acceptability. Therefore, unnecessary interventions should be minimized.

\begin{figure}[tb]
    \centering
    \includegraphics[width=0.9\linewidth]{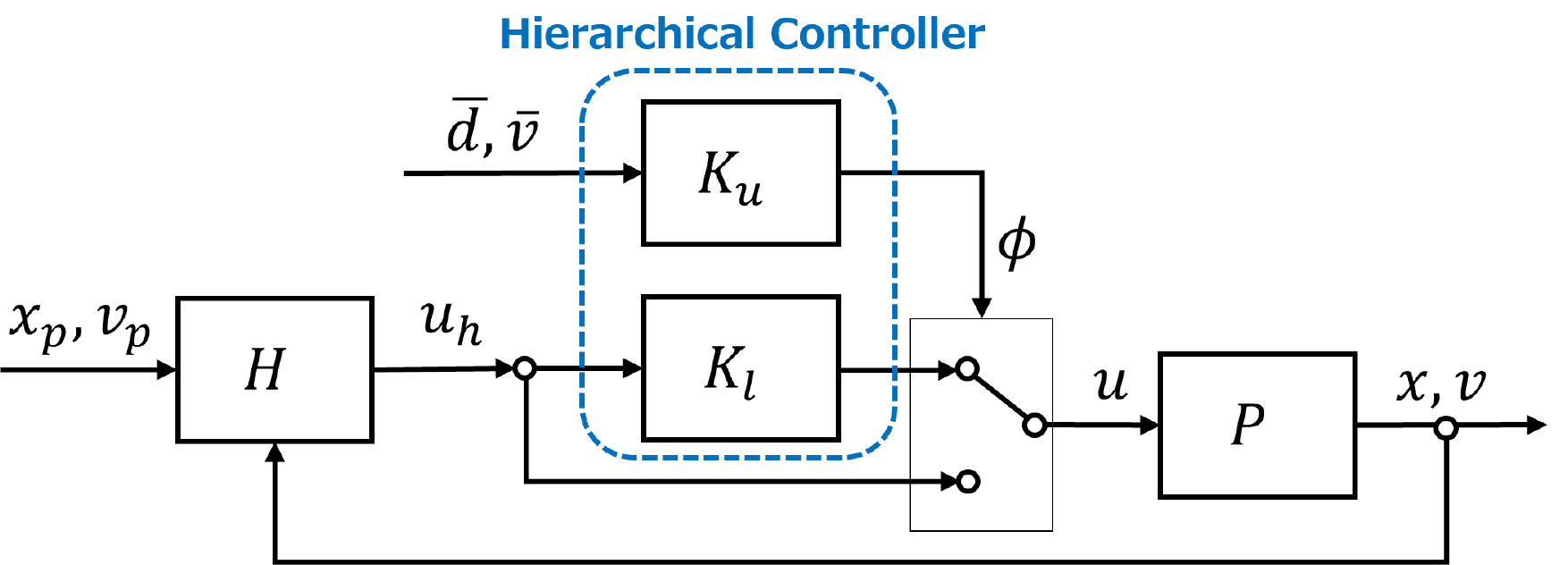}
    \caption{Overall structure of the proposed hierarchical control system}
    \label{fig:block}
\end{figure}

Accordingly, we aim to achieve traffic flow stabilization while suppressing unnecessary interventions as much as possible. To this end, we propose the hierarchical control system shown in Fig.~\ref{fig:block}.
In the figure, the driver $H$ performs accelerator and brake operations, and generates the corresponding acceleration input $u_h$ based on the position and velocity of the preceding vehicle, $x_p$ and $v_p$, as well as those of the controlled vehicle, $x$ and $v$. In addition, the average inter-vehicle distance $\bar{d}$ and average velocity $\bar{v}$ of the following vehicle platoon are obtained and used by the upper-layered controller $K_u$. Based on the traffic state of the following vehicle platoon, the upper-layered controller $K_u$ evaluates the influence of the controlled vehicles on traffic flow and generates an activation signal $\phi$. Each element of $\phi$ indicates whether the corresponding lower-layered controller should be activated.
The activated lower-layered controller $K_l$ intervenes in the acceleration input $u_h$ and generates the modified acceleration input $u$ to suppress traffic congestion. Subsequently, the modified acceleration input $u$ is applied to the vehicle $P$.
In the following subsections, we present the models of the vehicle $P$ and the driver $H$.

\subsection{Vehicle Model $P$}

The vehicle model $P$ is described by the discrete-time ballistic model developed by \cite{treiber2012traffic}. The longitudinal position $x$ and velocity $v$ of the vehicle are given by
\begin{align}
x(k+1)
&=
x(k)
+
\Delta t\,v(k)
+
\frac{1}{2}\Delta t^2 u(k),
\label{eq:dyn_pos}
\\
v(k+1)
&=
v(k)
+
\Delta t\,u(k),
\label{eq:dyn_vel}
\end{align}
where $\Delta t$ denotes the sampling period.

\subsection{Driver Model $H$}

The driver model $H$ is described by the Intelligent Driver Model (IDM) proposed by \cite{Treiber_2000}.
The IDM outputs the driver acceleration $u_h(k)$, which mimics the driver's car-following behavior, based on the inter-vehicle distance $d(k)=x_p(k)-x(k)$ and the relative velocity $\Delta v(k)=v(k)-v_p(k)$ with respect to the preceding vehicle.

The acceleration $u_h$ is generated by the IDM as follows:
\begin{equation}
u_h(k)
=
u_{\mathrm{IDM}}
\left[
1
-
\left(
\frac{v(k)}{v_{\mathrm{des}}}
\right)^{\delta}
-
\left(
\frac{d^{\ast}(k)}{d(k)}
\right)^2
\right],
\label{eq:idm_acc}
\end{equation}
where $v_{\mathrm{des}}$ denotes the desired velocity, $u_{\mathrm{IDM}}$ denotes the maximum acceleration, and $\delta$ is the acceleration exponent of the IDM. In addition, $d^{\ast}(k)$ denotes the desired inter-vehicle distance. In this manuscript, $d^{\ast}(k)$ is determined by the current velocity and the relative velocity with respect to the preceding vehicle as follows:
\begin{equation}
d^{\ast}(k)
=
d_0
+
T v(k)
+
\frac{v(k)\Delta v(k)}
{2\sqrt{u_{\mathrm{IDM}} b}},
\label{eq:idm_dxstar}
\end{equation}
where $b$ denotes the desired deceleration, $d_0$ denotes the jam distance, and $T$ denotes the safe time headway.

\section{Hierarchical Controller Design}

The proposed hierarchical control system is designed to achieve two primary objectives.  
The first objective is to suppress traffic congestion. Specifically, the proposed controller suppresses velocity fluctuations in traffic flow by regulating the acceleration and deceleration behaviors of the controlled vehicle so that unnecessary deceleration is not induced in the following vehicle platoon.
The second objective is to reduce driver burden by suppressing unnecessary interventions. To achieve the second objective, the upper-layered controller selectively activates the lower-layered controllers according to the traffic state. As a result, interventions are applied only when they are necessary for traffic congestion suppression.


\subsection{Upper-layered Controller $K_u$}
To simplify the notation, we focus on the case of a single controlled vehicle. Accordingly, the activation signal $\phi$ is treated as a scalar variable. Even in the case of multiple controlled vehicles, the following discussion applies by vectorizing $\phi$.

The upper-layered controller $K_u$ evaluates whether velocity fluctuations propagate through the following vehicle platoon and outputs $\phi$. Specifically, $\phi$ is determined based on the average inter-vehicle distance $\bar{d}$ and the average velocity $\bar{v}$ of the $n$ following vehicles.

The activation signal is given as follows:
\begin{equation}
\phi=
\begin{cases}
1, & g(\bar d,\bar v)>0,
\\[5pt]
0, & g(\bar d,\bar v)\leq 0,
\end{cases}
\label{eq:upper_F}
\end{equation}
where $\phi = 1$ indicates a request for intervention to activate the lower-layered controller $K_l$, whereas $\phi = 0$ indicates no such request. In addition, $g(\bar d,\bar v)$ is a function that evaluates whether the velocity reduction caused by deceleration of the preceding vehicle propagates to the following vehicle platoon.  
The function $g(\bar d,\bar v)$ is obtained through a data-driven approach. Specifically, it is obtained in the simulation preparation described in subsection~\ref{pre2}.
Accordingly, the upper-layered controller determines the activation signal $\phi$ based on the state of traffic flow. The lower-layered controller is activated when $\phi = 1$.

\subsection{Lower-layered Controller $K_l$}

The lower-layered controller $K_l$ intervenes in driver control actions to suppress traffic congestion.

Based on the driver acceleration input $u_h(k)$ and the states of the preceding and following vehicles, the lower-layered controller determines the intervention amount $u_a(k)$. The total acceleration input $u(k)$ is then generated as
\begin{equation}
u(k)=u_h(k)+u_a(k).
\label{eq:final_u}
\end{equation}

As studied by \cite{SAIFUZZAMAN2014379}, following vehicles decelerate when they perceive a risk of collision based on the inter-vehicle distance and relative velocity with respect to the preceding vehicle. 
In this manuscript, the propagation of a deceleration wave to following vehicles is determined based on a threshold of Time-to-Collision (TTC), which was introduced by \cite{hayward1972near}.

TTC is defined as follows:
\begin{equation}
\mathrm{TTC}(k)=
\begin{cases}
\dfrac{d(k)}{\Delta v(k)},
& \Delta v(k)>0,
\\[10pt]
\infty,
& \Delta v(k)\leq 0,
\end{cases}
\end{equation}
where $\Delta v(k)>0$ represents the situation in which the following vehicle is approaching the preceding vehicle. In this case, TTC decreases monotonically with respect to the relative velocity $\Delta v(k)$ and increases with respect to the inter-vehicle distance $d(k)$. TTC represents the remaining time until a collision occurs, assuming the current velocities remain constant. Therefore, a smaller TTC indicates a higher risk of collision and implicitly requires a larger braking effort from the following vehicle to ensure safety.

Each controlled vehicle must maintain two types of TTC: one with respect to the preceding vehicle, denoted by $T_p$, to ensure its own safety, and another with respect to the following vehicle, denoted by $T_f$, to suppress the propagation of velocity reductions.  To maintain sufficient TTC margins for both, we use the Control Barrier Function (CBF) framework. The barrier functions based on $T_p$ and $T_f$ are defined as follows:
\begin{align}
h_p(k)
&:=
d_p(k)
-
T_p \Delta v_p(k),
\label{eq:hf}
\\
h_f(k)
&:=
d_f(k)
-
T_f \Delta v_f(k),
\label{eq:hb}
\end{align}
where $h_p(k)>0$ indicates that TTC with respect to the preceding vehicle is maintained above $T_p$, thereby ensuring safety. Similarly, $h_f(k)>0$ indicates that TTC with respect to the following vehicle is maintained above $T_f$, thereby ensuring a sufficient temporal margin.

Using these barrier functions, we formulate the following optimization problem to determine the intervention amount $u_a$ that suppresses velocity reduction of following vehicles while ensuring safety with respect to the preceding vehicle.
\smallskip
\begin{Problem}
\label{prob:lower}
\begin{align}
\max_{u_a(k),\,T_b} \quad & T_b \\
\text{s.t.} \quad
&
(\ref{eq:dyn_pos}),\,
(\ref{eq:dyn_vel}),\,
(\ref{eq:final_u}),\,
(\ref{eq:hf}),\,
(\ref{eq:hb}),
\notag
\\
&
\frac{h_p(k+1)-h_p(k)}{\Delta t}
\ge
-\alpha h_p(k),
\\
&
\frac{h_f(k+1)-h_f(k)}{\Delta t}
\ge
-\alpha h_f(k),
\\
&
d_p(k)\ge d_{\min},
\\
&
-u_{\max}\le u(k)\le u_{\max},
\\
&
v_{\min}\le v(k)\le v_{\max}.
\end{align}
\end{Problem}

Through this optimization, the intervention amount $u_a(k)$ is obtained to maintain safety with respect to the preceding vehicle while maximizing the temporal margin with respect to the following vehicle. $u_a(k)$ is added to the driver acceleration $u_h(k)$ and applied to the vehicle as the final acceleration input.

\section{Experiment on Traffic Simulator}

In this section, we evaluate the effectiveness of the proposed hierarchical control system using a traffic simulator developed and calibrated with real-world data. In particular, we demonstrate that the upper-layered controller can macroscopically suppress traffic congestion without unnecessary interventions in human driving.

\subsection{Preparation 1: Simulator Development}
\begin{figure*}[t]
    \centering
    \includegraphics[width=0.9\textwidth]{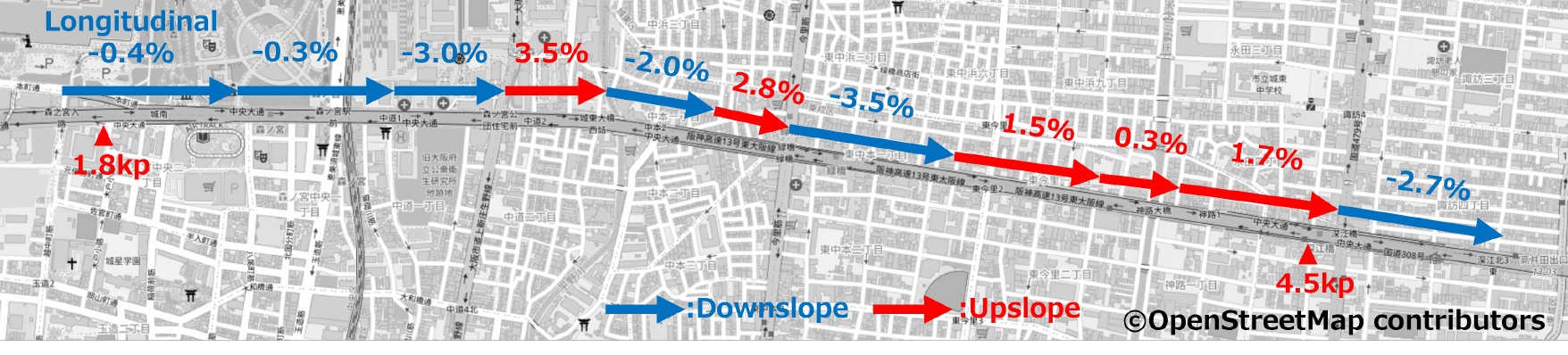}
    \caption{Hanshin Expressway Route 13 Higashi-Osaka Line near Morinomiya Junction (1.8--4.6 kp)\\
    Reference:https://zen-traffic-data.net/english/outline/dataprovision.html?area=morinomiya
    }
    \label{fig:ztd_area}
\end{figure*}

A traffic simulator is constructed using real-world traffic data.
The data used here are Zen Traffic Data (ZTD) provided by \cite{ZTD}. ZTD is a large-scale, high-resolution trajectory dataset obtained using image sensing technology and contains continuous trajectory information for almost all vehicles.
ZTD obtained on the Hanshin Expressway Route 13 Higashi-Osaka Line (toward Nara), near Morinomiya Junction (1.8--4.6 kp), are used. As shown in Fig.~\ref{fig:ztd_area}, highly accurate vehicle trajectory data with a temporal resolution of $0.1\,\mathrm{s}$ are available.
In addition, this manuscript uses one hour of traffic data, denoted as ``L003\_F001,'' collected from 10:00 to 11:00.

The simulation environment is built using the real-world data, and vehicle behavior is reproduced via equation~(\ref{eq:idm_acc}). To accurately match real-world driving, three adjustments were made to the IDM.

\begin{enumerate}
\item \textbf{Road Gradient:}
Because the road gradient changes throughout the target section, standard IDM acceleration alone is insufficient. To account for this, the simulation incorporates gravity effects based on the road gradient angle $\theta$ and gravitational acceleration $g$:
\begin{equation}
u_{\mathrm{use}} = u - g\sin\theta.
\end{equation}

\item \textbf{Perception Delay:}
To mimic actual human reaction times, the model introduces a $0.5\,\mathrm{s}$ perception delay.

\item \textbf{Vehicle Types:}
Since passenger cars and trucks exhibit different car-following behaviors, separate IDM parameters are applied to each vehicle type.
\end{enumerate}

The IDM parameters were calibrated to ensure that the simulation accurately reproduces the real-world dataset. Taking the parameter values from \cite{LIU2016706}, the parameters were adjusted to the values shown in Table~\ref{tab:idm_param} to reproduce the observed traffic flow.

\begin{table}[tb]
\centering
\caption{IDM parameters}
\label{tab:idm_param}
\begin{tabular}{lcc}
\hline
Parameter & Passenger car & Truck \\ \hline
$v_{\mathrm{des}}$ [m/s]
& 27.0
& 20.6
\\
$T$ [s]
& 1.19
& 1.76
\\
$d_0$ [m]
& 0.85
& 1.11
\\
$u_{\mathrm{IDM}}$ [m/s$^2$]
& 1.00
& 0.77
\\
$b$ [m/s$^2$]
& 2.26
& 1.71
\\
$\delta$
& 4
& 4
\\ \hline
\end{tabular}
\end{table}

\subsection{Preparation 2: Controller Settings}
\label{pre2}

\subsubsection{Settings of the Upper-layered Controller}

In the upper-layered controller, it is necessary to construct the function $g(\bar d,\bar v)$, which determines whether a deceleration wave propagates through the following vehicle platoon when the preceding vehicle decelerates.

\begin{figure}[tb]
    \centering
    \includegraphics[width=0.95\linewidth]{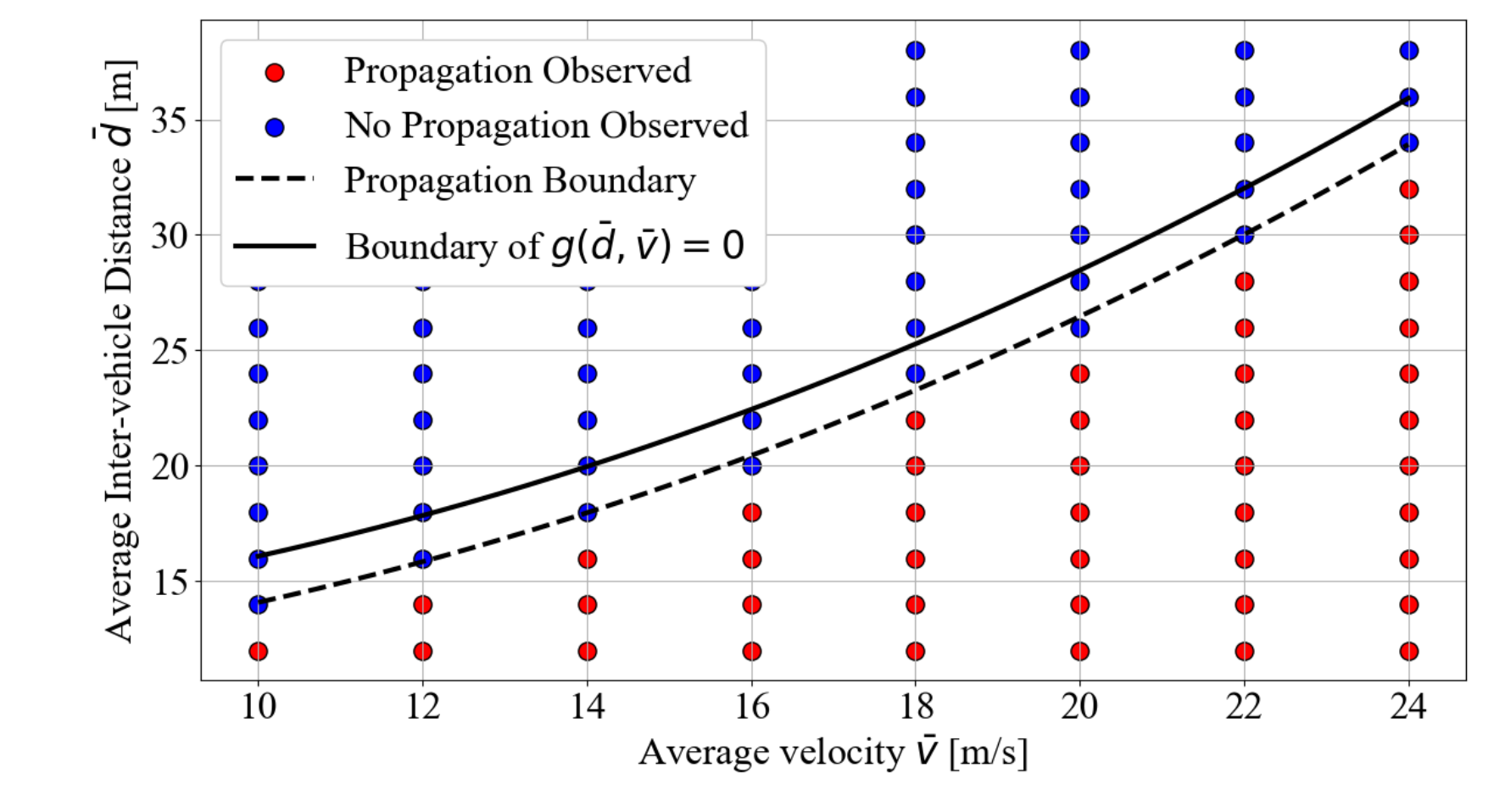}
    \caption{Simulation results of deceleration wave propagation and the resulting intervention boundary}
    \label{fig:upper_boundary}
\end{figure}

A data-driven approach is adopted for the design of the function $g(\bar d,\bar v)$.
In particular, data collection is performed assuming that vehicles initially travel according to the IDM with uniform inter-vehicle spacing and uniform velocity.
We then analyze the average inter-vehicle distance, average velocity, and whether the deceleration wave propagates, which indicates the likelihood of triggering traffic congestion. The function $g(\bar d,\bar v)$ is subsequently constructed from the determined threshold. The result is illustrated in Fig.~\ref{fig:upper_boundary}.
In the figure, the horizontal and vertical axes represent the average velocity $\bar v$ and the average inter-vehicle distance $\bar d$ of the following vehicle platoon at the initial time, respectively. The red points indicate that the deceleration wave is propagated, whereas the blue points indicate that no propagation occurs.

Function $g(\bar d,\bar v)=0$ is constructed from the boundary between these two regions. By incorporating a safety margin, the following function was obtained:
\begin{equation}
g(\bar d,\bar v)
=
0.04464\,\bar v^2
-
0.1012\,\bar v
+
12.63
-
\bar d
\label{eq:g_function}.
\end{equation}
In Fig.~\ref{fig:upper_boundary}, the dashed line represents the fitted propagation boundary, whereas the solid line represents the boundary defined by $g(\bar d,\bar v)=0$ after incorporating the safety margin.

\subsubsection{Settings of the Lower-layered Controller}

The parameters used in Problem~\ref{prob:lower} of the lower-layered controller are shown in Table~\ref{tab:sim_params}.
\begin{table}[t]
    \centering
    \caption{Parameters used in the lower-layered controller}
    \label{tab:sim_params}
    \begin{tabular}{c|c}
        \hline
        Parameter & Value \\ \hline
        $\Delta t$ & $0.1\,\mathrm{s}$ \\
        $u_{\max}$ & $5.0\,\mathrm{m/s^2}$ \\
        $d_{\min}$ & $2.0\,\mathrm{m}$ \\
        $v_{\min}$ & $0.0\,\mathrm{m/s}$ \\
        $v_{\max}$ & $39.0\,\mathrm{m/s}$ \\
        $\alpha$ & $0.8$ \\
        $T_f$ & $5.0\,\mathrm{s}$ \\ \hline
    \end{tabular}
\end{table}
In addition, IPOPT in CasADi developed by \cite{andersson2019casadi}, which is a nonlinear programming solver, was used to solve the optimization problem.

\subsection{Simulation Results}
In the traffic simulator, controlled vehicles were placed every five vehicles throughout the traffic stream.
Traffic flow simulations for 17 minutes were conducted for the following three cases:
\begin{description}
    \item[Case I:]
    Without intervention
    \item[Case I\hspace{-0.1em}I:]
    With the proposed hierarchical controller
    \item[Case I\hspace{-0.1em}I\hspace{-0.1em}I:]
    With the lower-layered controller ($\phi \equiv 1$)
\end{description}

The effectiveness of the proposed system was evaluated from the following two perspectives:

\begin{itemize}
    \item[(A)]
    Traffic congestion suppression:  
    This evaluation investigates whether velocity reduction and the propagation of deceleration waves in traffic flow are suppressed. 

    \item[(B)]
    Intervention efficiency:  
    This evaluation investigates how effectively unnecessary intervention to the driver can be reduced. 
\end{itemize}

\begin{figure}[tb]
    \centering
    \includegraphics[width=\linewidth]{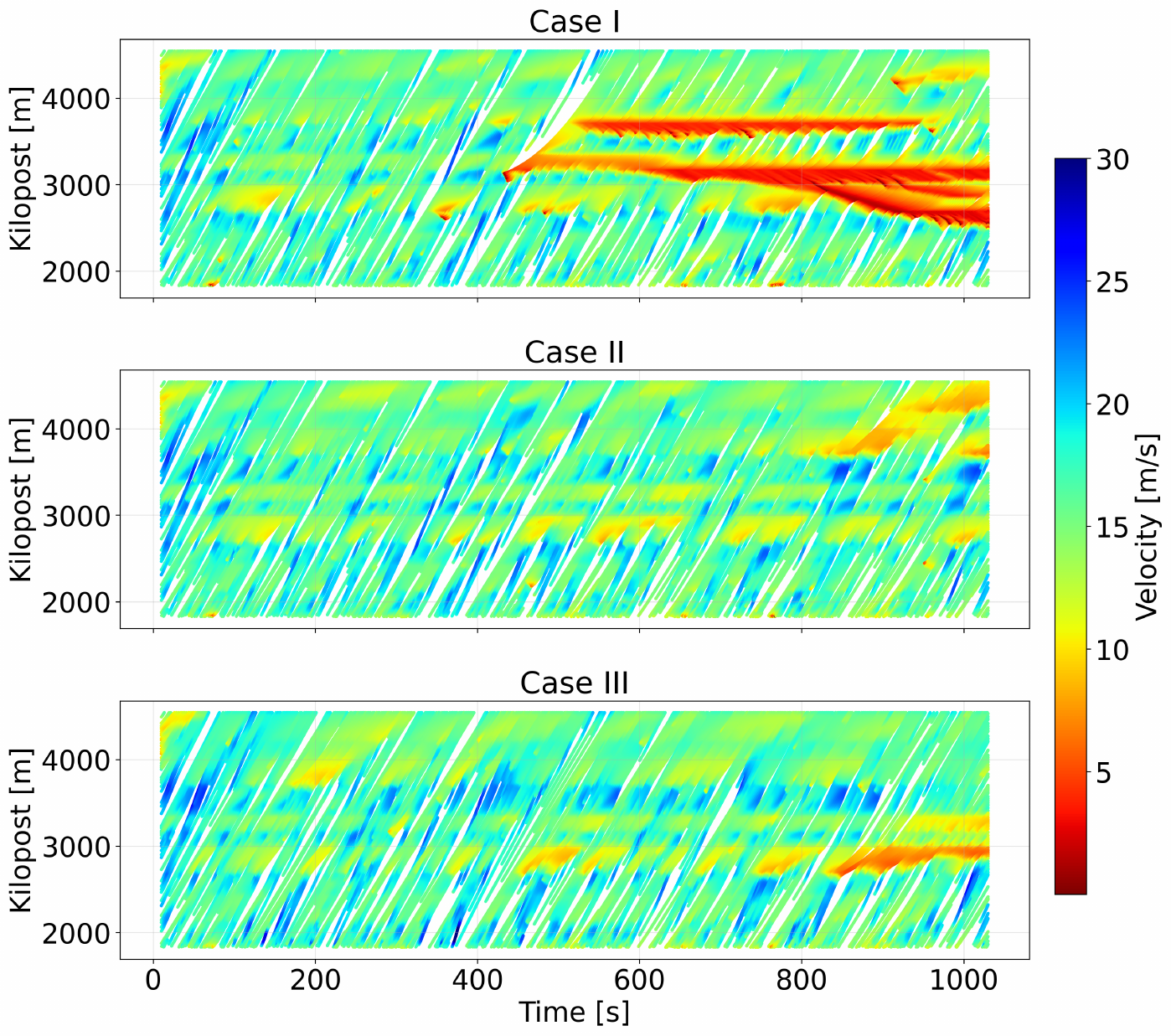}
    \caption{Spatiotemporal velocity distributions}
    \label{Spatiotempora}
\end{figure}

First, in order to evaluate (A) Traffic congestion suppression, the spatiotemporal velocity distributions under each condition were compared. The results are shown in Fig.~\ref{Spatiotempora}.
In the figure, a large low-velocity region appears in Case I around 400--1000\,s, indicating the occurrence of traffic congestion. In contrast, the low-velocity region is significantly reduced in Cases I\hspace{-0.1em}I and I\hspace{-0.1em}I\hspace{-0.1em}I.

\begin{figure}[tb]
    \centering
    \includegraphics[width=0.9\linewidth]{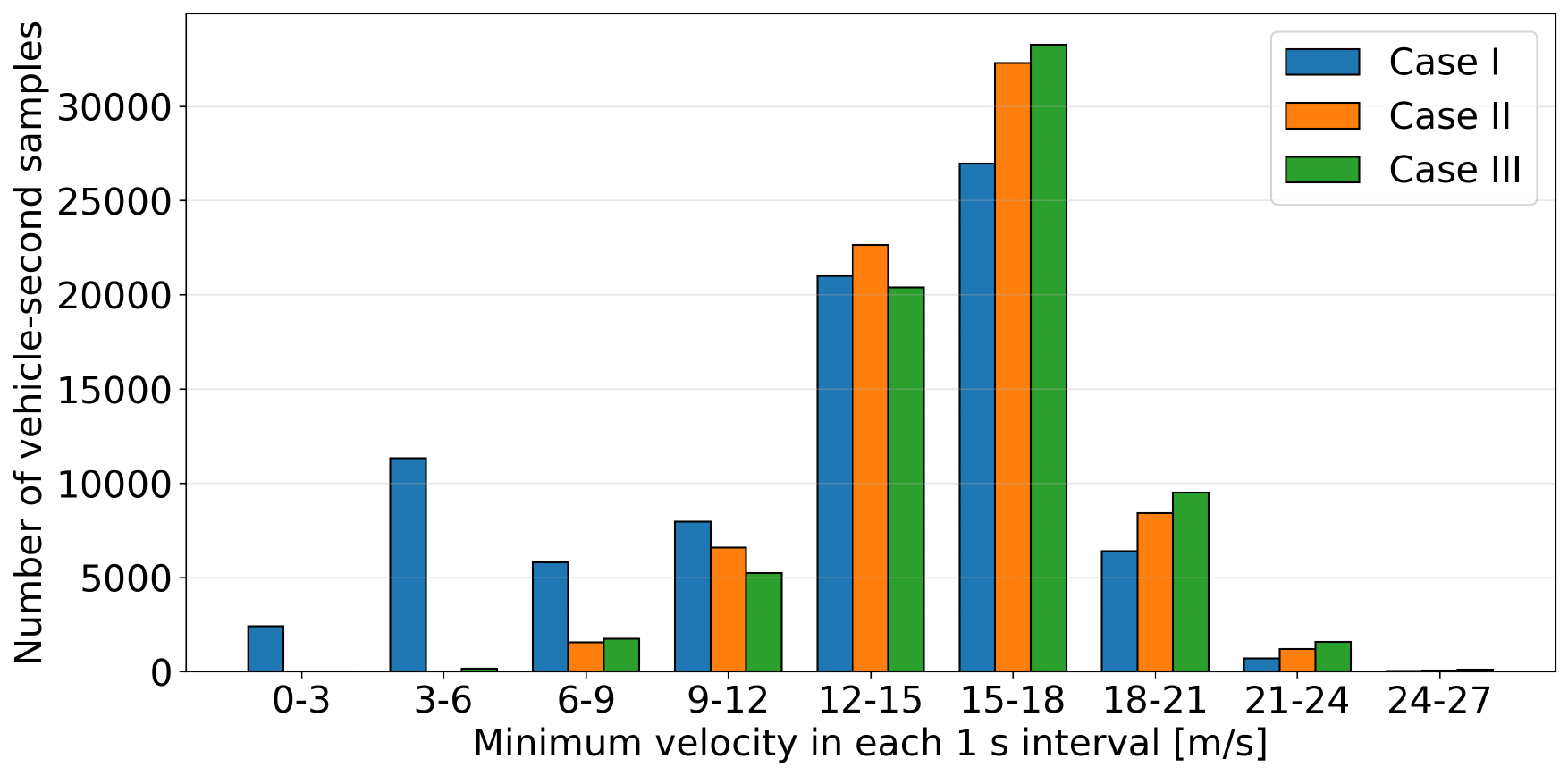}
    \caption{Distribution of minimum velocity in each 1 s interval}
    \label{fig:velocity_histogram}
\end{figure}

To further evaluate the traffic state from a statistical perspective, the distribution of minimum velocity in each $1\,\mathrm{s}$ interval for all vehicles was compared. The results are shown in Fig.~\ref{fig:velocity_histogram}.
In the figure, the proposed TTC-based control in Cases I\hspace{-0.1em}I and I\hspace{-0.1em}I\hspace{-0.1em}I successfully shifts the entire velocity distribution toward a higher range. In particular, there is a significant reduction in the number of vehicles with velocities below 12 m/s, compared with Case I. These results indicate that the proposed control system efficiently suppresses traffic congestion.

Furthermore, we investigate the effects of the upper-layered controller $K_u$.
There is essentially no major difference between Case I\hspace{-0.1em}I, where $K_u$ is introduced to moderately inactivate the lower-layered controller $K_l$, and Case I\hspace{-0.1em}I\hspace{-0.1em}I, where $K_l$ remains constantly active.
However, in the velocity range above $15\,\mathrm{m/s}$, it is apparent that velocities in Case I\hspace{-0.1em}I are slightly lower than those in Case I\hspace{-0.1em}I\hspace{-0.1em}I.

The average velocity under each condition is shown in Table~\ref{tab:mean_velocity}.
\begin{table}[tb]
\centering
\caption{Average velocity under each condition}
\label{tab:mean_velocity}
\begin{tabular}{lc}
\hline
Condition & Average velocity [m/s] \\ \hline
Case I & 12.789 \\
Case I\hspace{-0.1em}I & 15.577 \\
Case I\hspace{-0.1em}I\hspace{-0.1em}I & 15.377 \\ \hline
\end{tabular}
\end{table}
From these results, both Cases I\hspace{-0.1em}I and I\hspace{-0.1em}I\hspace{-0.1em}I improve the average velocity compared with Case I. Since there is no major difference between Cases I\hspace{-0.1em}I and I\hspace{-0.1em}I\hspace{-0.1em}I, we conclude that $K_u$ successfully suppresses unnecessary interventions.

To evaluate (B) intervention efficiency, the total intervention time for controlled vehicles was compared between Case I\hspace{-0.1em}I and Case I\hspace{-0.1em}I\hspace{-0.1em}I. The total intervention time was $10071.3\,\mathrm{s}$ for Case I\hspace{-0.1em}I\hspace{-0.1em}I, whereas it was reduced to $2369.1\,\mathrm{s}$ for Case I\hspace{-0.1em}I. Notably, Case I\hspace{-0.1em}I reduces the frequency of intervention by $76.5\%$ compared to Case I\hspace{-0.1em}I\hspace{-0.1em}I. 
This reduction in intervention frequency is expected to reduce driver frustration and anxiety, thereby enhancing user acceptance.

While there is no major difference in traffic congestion suppression performance between the two cases, Case I\hspace{-0.1em}I outperforms Case I\hspace{-0.1em}I\hspace{-0.1em}I in terms of intervention frequency. These results demonstrate that the hierarchical control in Case I\hspace{-0.1em}I contributes to suppressing traffic congestion without imposing a significant burden on the driver.

\section{Conclusion}

In this manuscript, a hierarchical control system was proposed to simultaneously suppress traffic congestion and reduce driver burden. In the proposed system, the lower-layered controller intervenes in microscopic driving behavior based on CBF to suppress the propagation of deceleration waves by maintaining sufficient temporal margins for following vehicles. Meanwhile, the upper-layered controller monitors the macroscopic traffic conditions of the following vehicle platoon and activates the lower-layered controllers only when the propagation of a deceleration wave is predicted.
Furthermore, the proposed system was evaluated using a traffic flow simulator. The simulation results demonstrated that the proposed system reduces intervention time by approximately $76.5\%$ while maintaining traffic congestion suppression performance. These results indicate that the proposed system has the potential to simultaneously stabilize traffic flow and reduce driver burden through the control of only a limited number of vehicles.

As future work, the upper-layered controller will be extended to predict future deceleration of the preceding vehicles rather than relying solely on the states of the following vehicle platoon. We will also introduce anticipatory interventions before deceleration waves emerge. Such a predictive framework is expected to achieve earlier and more effective traffic flow stabilization.

\bibliography{ifacconf}             

@Article{su13073932,
AUTHOR = {Paiva, Sara and Pañeda, Xabiel García and Corcoba, Victor and García, Roberto and Morán, Próspero and Pozueco, Laura and Valdés, Marina and del Camino, Covadonga},
TITLE = {User Preferences in the Design of Advanced Driver Assistance Systems},
JOURNAL = {Sustainability},
VOLUME = {13},
YEAR = {2021},
NUMBER = {7},
ARTICLE-NUMBER = {3932},
ISSN = {2071-1050},
DOI = {10.3390/su13073932}
}

@ARTICLE{7490340,
  author={Paden, Brian and Čáp, Michal and Yong, Sze Zheng and Yershov, Dmitry and Frazzoli, Emilio},
  journal={IEEE Transactions on Intelligent Vehicles}, 
  title = {A survey of motion planning and control techniques for self-driving urban vehicles},
  year={2016},
  volume={1},
  number={1},
  pages={33-55},
  doi={10.1109/TIV.2016.2578706}}

@article{SAIFUZZAMAN2014379,
title = {Incorporating human-factors in car-following models: A review of recent developments and research needs},
journal = {Transportation Research Part C: Emerging Technologies},
volume = {48},
pages = {379-403},
year = {2014},
issn = {0968-090X},
doi = {https://doi.org/10.1016/j.trc.2014.09.008},
author = {Mohammad Saifuzzaman and Zuduo Zheng}
}

@book{treiber2012traffic,
  title={Traffic flow dynamics: data, models and simulation},
  author={Treiber, Martin and Kesting, Arne},
  year={2012},
  publisher={Springer Science \& Business Media}
}

@ARTICLE{9928340,
  author={Ferrara, Antonella and Incremona, Gian Paolo and Birliba, Eugeniu and Goatin, Paola},
  journal={IEEE Open Journal of Intelligent Transportation Systems}, 
  title={Multi-scale model-based hierarchical control of freeway traffic via platoons of connected and automated vehicles}, 
  year={2022},
  volume={3},
  number={},
  pages={799-812},
  doi={10.1109/OJITS.2022.3217001}}

@article{Zhao_2023,
   title={Safety-critical traffic control by connected automated vehicles},
   volume={154},
   ISSN={0968-090X},
   DOI={10.1016/j.trc.2023.104230},
   journal={Transportation Research Part C: Emerging Technologies},
   publisher={Elsevier BV},
   author={Zhao, Chenguang and Yu, Huan and Molnar, Tamas G.},
   year={2023},
   month=sep, pages={104230} }

@INPROCEEDINGS{8796030,
  author={Ames, Aaron D. and Coogan, Samuel and Egerstedt, Magnus and Notomista, Gennaro and Sreenath, Koushil and Tabuada, Paulo},
  booktitle={2019 18th European Control Conference (ECC)}, 
  title={Control barrier functions: Theory and applications}, 
  year={2019},
  volume={},
  number={},
  pages={3420-3431},
  doi={10.23919/ECC.2019.8796030}}

@article{Sugiyama_2008,
doi = {10.1088/1367-2630/10/3/033001},
year = {2008},
month = {mar},
publisher = {},
volume = {10},
number = {3},
pages = {033001},
author = {Sugiyama, Yuki and Fukui, Minoru and Kikuchi, Macoto and Hasebe, Katsuya and Nakayama, Akihiro and Nishinari, Katsuhiro and Tadaki, Shin-ichi and Yukawa, Satoshi},
title = {Traffic jams without bottlenecks—experimental evidence for the physical mechanism of the formation of a jam},
journal = {New Journal of Physics}
}

@ARTICLE{9037275,
  author={Li, Tingting and Wu, Jianping and Chan, Ching-Yao and Liu, Mingyu and Zhu, Chunli and Lu, Weixin and Hu, Kezhen},
  journal={IEEE Access}, 
  title={A Cooperative Lane Change Model for Connected and Automated Vehicles}, 
  year={2020},
  volume={8},
  number={},
  pages={54940-54951},
  doi={10.1109/ACCESS.2020.2981169}}

@article{TALEBPOUR2016143,
title = {Influence of connected and autonomous vehicles on traffic flow stability and throughput},
journal = {Transportation Research Part C: Emerging Technologies},
volume = {71},
pages = {143-163},
year = {2016},
issn = {0968-090X},
doi = {https://doi.org/10.1016/j.trc.2016.07.007},
author = {Alireza Talebpour and Hani S. Mahmassani},
}

@article{Treiber_2000,
   title={Congested traffic states in empirical observations and microscopic simulations},
   volume={62},
   ISSN={1095-3787},
   DOI={10.1103/physreve.62.1805},
   number={2},
   journal={Physical Review E},
   publisher={American Physical Society (APS)},
   author={Treiber, Martin and Hennecke, Ansgar and Helbing, Dirk},
   year={2000},
   month=aug, pages={1805–1824} }

@article{hayward1972near,
  title={Near miss determination through use of a scale of danger},
  author={Hayward, John C},
  year={1972},
  publisher={Pennsylvania State University University Park}
}

@article{andersson2019casadi,
  title     = {CasADi: a software framework for nonlinear optimization and optimal control},
  author    = {Andersson, Joel A. E. and Gillis, Joris and Horn, Greg and Rawlings, James B. and Diehl, Moritz},
  journal   = {Mathematical Programming Computation},
  volume    = {11},
  number    = {1},
  pages     = {1--36},
  year      = {2019},
  publisher = {Springer},
}

@article{LIU2016706,
title = {Modeling and simulation of the car-truck heterogeneous traffic flow based on a nonlinear car-following model},
journal = {Applied Mathematics and Computation},
volume = {273},
pages = {706-717},
year = {2016},
issn = {0096-3003},
doi = {https://doi.org/10.1016/j.amc.2015.10.032},
author = {Lan Liu and Liling Zhu and Da Yang}
}

@ARTICLE{6515636,
  author={Ploeg, Jeroen and van de Wouw, Nathan and Nijmeijer, Henk},
  journal={IEEE Transactions on Control Systems Technology}, 
  title={Lp String Stability of Cascaded Systems: Application to Vehicle Platooning}, 
  year={2014},
  volume={22},
  number={2},
  pages={786-793},
  doi={10.1109/TCST.2013.2258346}}

@INPROCEEDINGS{531196,
  author={Swaroop, D. and Hedrick, J.K.},
  booktitle={Proceedings of 1995 American Control Conference - ACC'95}, 
  title={String stability of interconnected systems}, 
  year={1995},
  volume={3},
  number={},
  pages={1806-1810 vol.3},
  doi={10.1109/ACC.1995.531196}}

@article{li2025safety,
  title={Safety-Critical Control of Connected Vehicle Systems Based on Barrier Functions},
  author={Li, Zeming and Liu, Yonggui and Shen, Zhiping},
  journal={IEEE Transactions on Intelligent Transportation Systems},
  year={2025},
  publisher={IEEE}
}

@article{FAN2025100188,
title = {Integrating micro and macro traffic control for mixed autonomy traffic},
journal = {Communications in Transportation Research},
volume = {5},
pages = {100188},
year = {2025},
issn = {2772-4247},
doi = {https://doi.org/10.1016/j.commtr.2025.100188},
author = {Tingting Fan and Jieming Chen and Edward Chung},
}

@misc{ZTD,
  author = {{Hanshin Expressway Co., Ltd.}},
  title = {Zen traffic data},
  year = {2018},
  note = {Available from: \url{https://zen-traffic-data.net/english/}. Accessed 21 May 2026}
}
                                                   







\end{document}